\documentclass{article}
\usepackage{ijcai26}

\usepackage{times}
\usepackage{soul}
\usepackage{url}
\usepackage[hidelinks]{hyperref}
\usepackage[utf8]{inputenc}
\usepackage[small]{caption}
\usepackage{graphicx}
\usepackage{amsmath}
\usepackage{amsthm}
\usepackage{booktabs}
\usepackage{algorithm}
\usepackage{algorithmic}
\usepackage[switch]{lineno}
\usepackage{subcaption}
\usepackage{float}

\title{Communicate-Predict-Act:
Evaluating Social Intelligence of Agents}

\author{
David Shoresh$^{1,2}$
\and
Sarit Kraus$^{4}$\and
Yonatan Loewenstein$^{1,2,3}$
\\
\affiliations
$^1$Edmond and Lily Safra Center for Brain Sciences\\
$^2$The Federmann Center for the Study of Rationality\\
$^3$The Alexander Silberman Institute of Life Science and the Department of Cognitive and Brain Sciences, The Hebrew University of Jerusalem\\
$^4$Department of Computer Science, Bar-Ilan University\\
}

\begin{document}

\maketitle

\begin{abstract}
As large language model (LLM) agents become more prevalent in real-world social settings, social intelligence will play an increasingly critical role. But social intelligence is still a poorly defined construct, for humans and artificial agents. We introduce a multiplayer arena of mixed cooperative–competitive social games to study LLM social intelligence. The controllability of LLM-based agents enables systematic evaluation, which also supports broader inferences about social intelligence per se. We evaluated eight diverse LLMs (24B–1T parameters) using a Communicate–Predict–Act (COMPACT) interaction protocol and fine-grained probing of social dynamics. Elo-style ratings reveal consistent performance differences across models, but this scalar measure provides only a partial characterization of social intelligence. To address this limitation, we analyze gameplay traces to extract socio-cognitive metrics capturing action prediction, communicative influence, strategic reasoning, and trade-offs under conflicting interests.
These socio-cognitive metrics exhibit strong intra-model consistency and they reliably predict pairwise agent advantage in game outcomes (AUC-ROC = 0.82). Feature-importance analysis indicates that surprisingly, influence, transparency, and adaptability are more predictive of success than Theory-of-Mind inference or deep planning. Together, our results advance a testable, multi-dimensional conception of social intelligence and provide empirical insights into the capacities that underpin it.
\end{abstract}

\section{Introduction}

AI agents are poised to take on roles in the real world, interacting with third parties, including other people and other agents, requiring them to navigate complex social situations. This creates a need for evaluating and training agents for social intelligence.

However, social intelligence is not well understood, even in humans. Various theories have attempted to define social intelligence bottom-up as a composite of different cognitive faculties\cite{SocialIntelligence,SocialSkills}. Components relating to social intelligence that have been proposed include understanding others ("Theory of Mind")\cite{SocialCognition}, influencing others\cite{Influence}, susceptibility to influence by others\cite{Susceptibility}, communication skills\cite{Communication}, strategic reasoning\cite{StrategicReasoning}, social learning\cite{SocialLearning} and conflict management\cite{Conflict}. Such decompositions can be intuitive, but they lack a unifying principle, and it is not clear how they relate to actual social performance.

A second problem is measurement. Social intelligence is exhibited in dynamic interactive contexts, but this is expensive to carry out at scale in a controlled manner with humans. Furthermore, humans aren't stationary, and any attempt to measure individual components of social intelligence can change their behavior. This complicates the ability to isolate the determinants of social intelligence.

For the first problem, we take a top-down approach, using relative performance in diverse social games across diverse opponents as a proxy for social intelligence and then search for the attributes that support it. As for the measurement problem, the advent of LLM agents, which can be deployed at scale and do not have memory beyond the context defined for them, provides new research opportunities. 

We propose a framework in which at each round of a game, agents must first communicate with each other in natural language, then expressly anticipate each others' actions, and then reason about their own actions and make decisions (communicate-predict-act, or "COMPACT"). This protocol generates rich socio-cognitive data, which allows us to deepen our investigation into the "social profile" of the agents and the determinants of their social intelligence. 

Our research questions can be summarized as follows:
\begin{description}
    
    \item[RQ1] To what extent can LLM-agents be rated on a single-dimensional scale of social intelligence?
    \item[RQ2] Do LLM agents exhibit self-consistent socio-cognitive attributes across diverse game settings?
    \item[RQ3] Which socio-cognitive attributes contribute most to general social intelligence?
\end{description}

\section{RQ1: A Social Intelligence Scalar}
\subsection{Defining Social Intelligence}

The "reward is enough" hypothesis states that "reward is enough to drive behaviour that exhibits abilities studied in natural and artificial intelligence, including ... \textit{social intelligence}..."\cite{RewardisEnough} (emphasis added). Indeed, even seemingly altruistic pro-social behavior that might seem unrelated to reward maximization, such as honesty or cooperation, are known to emerge from reward-driven policies through the reciprocal and temporal dynamics of games\cite{Axelrod}. Note that reward-maximization is distinct from the Nash equilibrium, since opponents may be boundedly rational, and a player with high social intelligence should be able to recognize and best-respond to that as well\cite{RewardisEnough}.

Drawing inspiration from the "reward is enough" hypothesis, we adopt an operative definition of social intelligence as the ability to achieve objectives in mixed cooperative-competitive multi-agent environments of variable sizes and compositions, while communicating and interacting with other agents over time. This is in accordance with characteristics of human social settings which usually have mixed competitive-cooperative reward structures, extensible number of participants, temporal dynamics and crucially, the opportunity to communicate. Research environments rarely possess all of these features.

It is not entirely clear that a single scalar social intelligence metric can differentiate LLM agents meaningfully in such a setup. For example, it might be that different agents have relative advantages in different social settings, or that they have relative advantages to each other (i.e., maybe social intelligence is non-transitive), or that LLM agents are simply too prompt-sensitive for consistent behavior. So, we would need to check for generalized differences across games and opponent compositions.

\subsection{Formalizing a Social Intelligence Scalar}
We assume a set of agents $A=\{A_i \}, i \in \{1, 2, ...\}$ play each other in a diverse set of social games $G=\{G_k\}, k \in \{1, 2, .. .\}$, in diverse compositions of participants. Each game match involving a particular subset of agents in a particular game is an event $e \in E$. Under the rules of the games (described later), each such event produces a reward for each of the participants, $R_{i, e}, i \in A, e \in E$, which depends on their actions. To quantify performance, we first adopt the probabilistic view envisaged in the Bradley-Terry model. This model assumes that each agent has a single skill metric measured by a scalar, $w_i, i \in A$. The probability that an agent $A_i$ will get a higher reward than $A_j$ in any given game match is modeled as:

\begin{equation}
    \log[P_{i, j \sim{A}, e\sim{E}}(R_{i, e} > R_{j, e})] \propto w_i - w_j 
    \label{single_w}
\end{equation}

The scalar parameters per player $i$, $w_i$, are measures of the skills of the agents. They are estimated with maximum likelihood estimation (MLE)\cite{MLE} from pairwise comparisons of agent rewards in the games using gradient descent. If the fixed player ratings $w_i$ can predict outcomes in games better than chance, this means that some inherent single social intelligence factor explains at least some of the variance in performance between the agents. This is the principle on which Elo scores are based\cite{Elo,BradleyTerry}. 

However, unlike most cases in which Elo is implemented, we have multiple game types. The contribution of social intelligence may vary between games. Therefore, it is useful to estimate $w_i$ together with a per-game bias:

\begin{equation}
    w_{i,k} = \mu_i + \delta_{i,k} \,
    \label{elo2}
\end{equation}

Where $\mu_i$ is a global rating for player $i$ and $\delta_{i,k}$ is a deviation for that player in game type $G_k$, allowing for some variability in agents' abilities across games. As long as $\delta_{i,k}$ does not completely shuffle the rankings of agents across games, this model could represent a single basic social intelligence with some flexibility surrounding player propensities in certain game types.  

Alternatively, it might be that performance of agents are better explained by multiple attributes, with different games being sensitive to different attributes. This would imply that there is not a single intelligence factor inherent to agents, only relative advantages that manifest differently in different games. To check for this, we can model a \textit{vector} of agent attributes $\boldsymbol{w_i}$, each of which are scaled differently by per-game vectors $\boldsymbol{g_k}$:

\begin{equation}
    \log[P_{i, j \sim{A}, e\sim{E}}(R_{i, e} > R_{j, e})] \propto (\boldsymbol{w_i} - \boldsymbol{w_j}) \cdot \boldsymbol{g_k}
    \label{w_g}
\end{equation}

The agent and game vectors in this model can also be estimated by MLE. If it significantly outperforms the single scalar model in Equation \ref{single_w}, for some dimension-size of the vectors $\boldsymbol{w}, \boldsymbol{g}$, then it could be said that the features determining agents' performances are too complex to be summarized by a single social intelligence scalar.

\subsection{Operationalizing Social Intelligence}
We propose an arena of mixed cooperative-competitive games\cite{MixedMotive}, with varying numbers of participants\cite{GroupSize}, in-game communication\cite{Strategic-Communication} and temporal dynamics (repeated and/or extensive form), possibly with partial information\cite{BayesianGames}. We argue that these characteristics are important for simulating human-like social interactions, which are not adequately captured by overly simplistic models\cite{SocialGames}. 

In our environment, at each round of the game, players engage in three stages: communicate, predict and act. The communication stage is conducted pairwise between players with limited message-passing in natural language. The prediction stage requires the players to privately reflect on the conversations and history of the game and predict the next actions of the other players. Players then simultaneously apply reasoning and make a decision.

We chose five games inspired from the literature, designed to test diverse social situations and challenges, spanning the competitive-cooperative spectrum:

\begin{enumerate}
    \item \underline{Coalition} - a classic cooperative game\cite{Coalition} where a subset of players form a coalition, dependent on an agreement regarding the distribution of the benefits. The game tests group formation and resource allocation;
    \item \underline{Scheduler} - a multiplayer generalization of "Battle of the Sexes", where players seek agreement on a choice from a set of options with different preference orders, but failure to coordinate is the worst case for all. The game tests coordination and compromise under preference conflict;
    \item \underline{Tragedy of Commons} - a classic public resource utilization game\cite{TragedyofCommons}. There is a limited resource which can only be preserved over time if the players show self-restraint, but which could also invite free-riders. The game tests social dilemmas between the individual and the collective good;
    \item \underline{Survivor} - a simplified version of "Diplomacy"\cite{KrausDiplomacy,Diplomacy} (or "Risk" \cite{zuckerman2009mixing}) but with no board. Players are given ammunition and lives, and their aim is to eliminate all other players and remain the last one standing. The game tests alliance formation and betrayal; 
    \item \underline{HUPI}\cite{Lupi} - a game where players try to pick the "Highest Unique Positive Integer", testing complex k-level reasoning\cite{CognitiveHierarchy}.
\end{enumerate}

Players were initialized with random names and private preferences, but the game logic is deterministic in all games, and the informational and reward structures are strictly symmetric. For example, in Survivor all players received the same amount of lives and ammunition but none were informed how much the others had; in Coalition all players had equal and transparent assets that they could contribute to a coalition; and in Scheduler the preference orders of players were arranged as a circulant matrix. Each game was run with two alternative semantic back-stories to test for agent sensitivity to framing. For example, Survivor is played either as a battle between cowboys or between pirates. Detailed game descriptions are provided in the supplementary material.

A pool of 8 diverse LLM agents were evaluated on the above games using the COMPACT methodology: GPT-5, Qwen3-235B A22B, Kimi-K2 (v0711), GPT-OSS 120B, Gemini-2.5-Flash-Lite, GPT-OSS 20B, Qwen3-30B, Mistral Small3-24B. Games were played in 2, 3, 4 and 5 player versions, producing 5x4=20 game types. Altogether 928 games were run, comprising almost half of all possible player-game-framing combinations. 

\begin{figure}
    \centering
    \begin{subfigure}{0.3\textwidth}
        \includegraphics[width=0.9\linewidth]{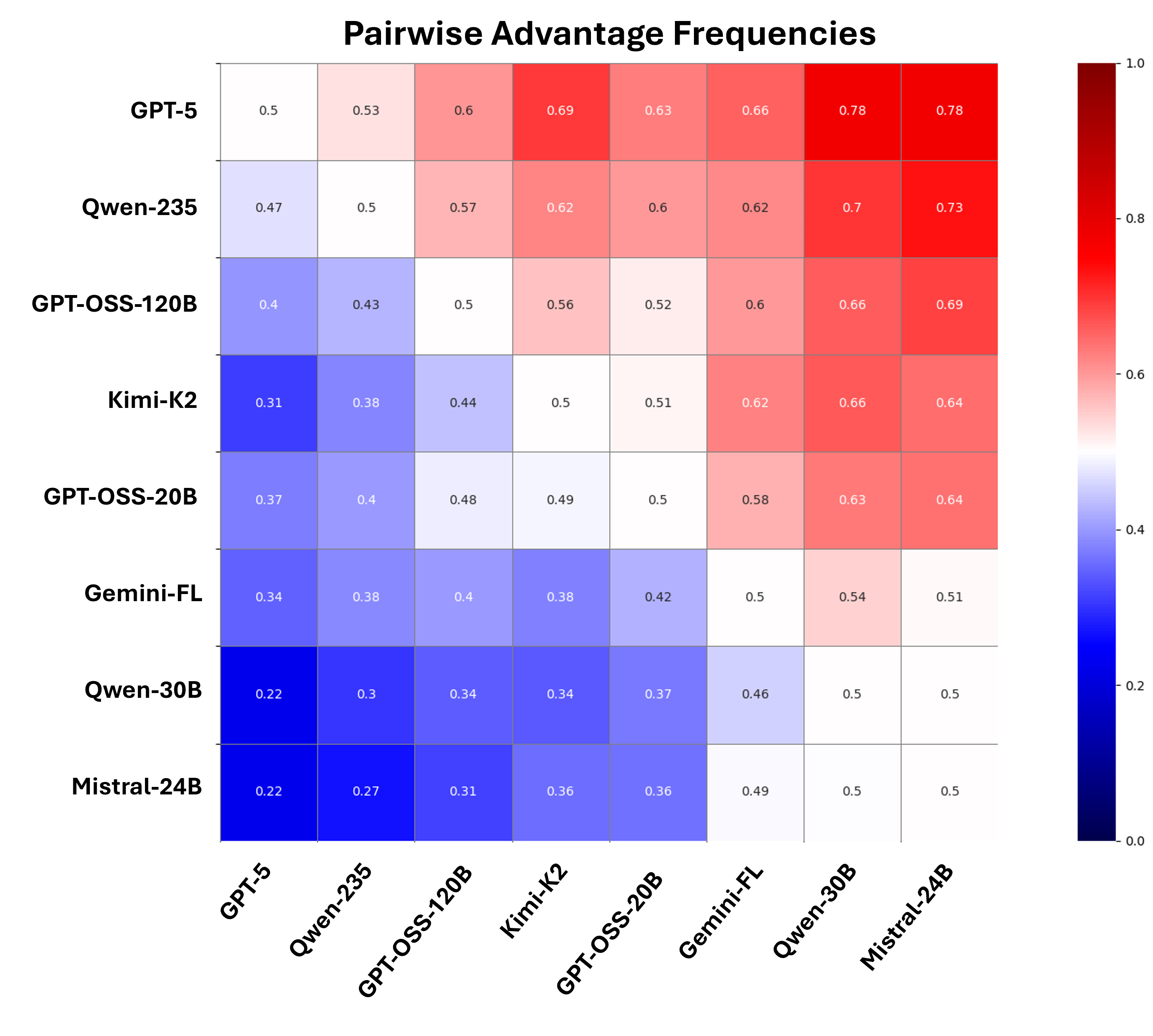}
        \caption{Games with communication}
        \end{subfigure}
    \begin{subfigure}{0.3\textwidth}
        \includegraphics[width=0.9\linewidth]{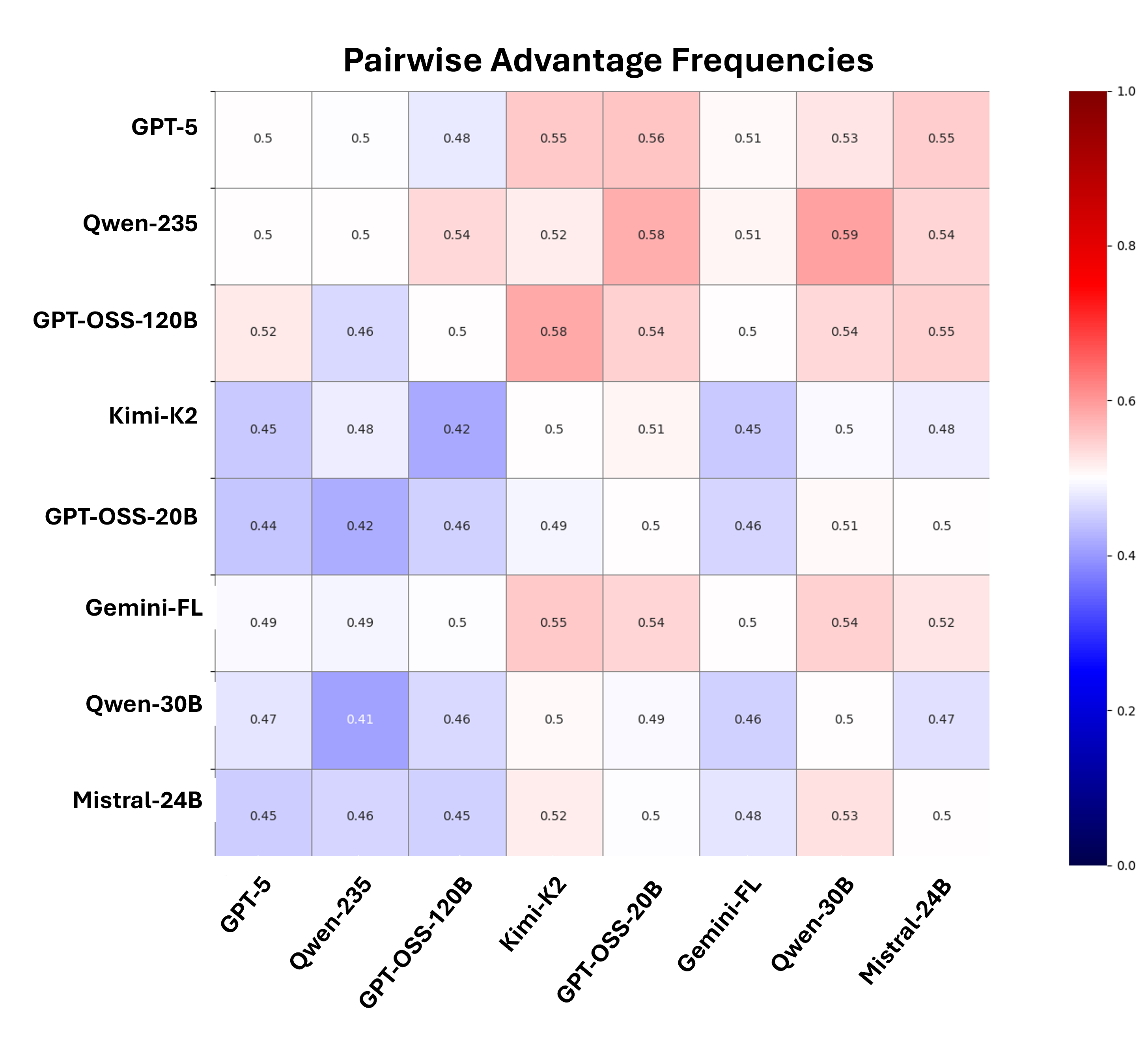}
        \caption{Games without communication}
    \end{subfigure}
    \caption{Heatmaps showing probabilities of agents outperforming each other.}
    \label{heatmaps}
\end{figure}

We also ran an ablation study in which the same games were played without communication. Heatmaps showing the probabilities of each player (rows) doing better than each other player (columns) globally, are shown in figure~\ref{heatmaps}. The top shows results for games with communication and the bottom without communication. In the games without communication, differences in performances have much diminished effect. This strengthens our confidence that the games are capturing aspects of social intelligence, and not only, say, analytic skills, and underlies the crucial importance of communication in social games.

\subsection{Elo Results}

\begin{table*}[t]
    \centering
    \begin{tabular}{ccccccc}
    \hline
    Model        & \textbf{Global} & Survivor & Scheduler & Coalition & Tragedy & HUPI \\
    \hline
    GPT-5        & \textbf{1603}   & 1566     & 1539      & 1680      & 1686    & 1647 \\
    Qwen3-235B   & \textbf{1551}   & 1479     & 1645      & 1516      & 1594    & 1573 \\
    GPT-OSS 120B & \textbf{1523}   & 1477     & 1541      & 1556      & 1529    & 1536 \\
    Kimi-K2      & \textbf{1514}   & 1536     & 1556      & 1588      & 1462    & 1443 \\
    GPT-OSS 20B  & \textbf{1488}   & 1424     & 1474      & 1479      & 1469    & 1581 \\
    Gemini-FL    & \textbf{1462}   & 1446     & 1472      & 1479      & 1472    & 1403 \\
    Qwen3-30B    & \textbf{1435}   & 1423     & 1403      & 1411      & 1464    & 1410 \\
    Mistral 24B  & \textbf{1420}   & 1396     & 1459      & 1409      & 1396    & 1362 \\
    \hline
    \end{tabular}
       \caption{ELO Scores}
        \label{elo-results}
\end{table*}

For the single skill rating model in Equation \ref{single_w}, we followed the classic Elo method\cite{Elo}, setting a 1500 prior for all ratings and a temperature of 400 in the logistic function:

\begin{equation}
    P(i \text{ beats } j) = \frac{1}{1 + 10^{(w_j - w_i)/400}} 
    \label{elo1}
\end{equation}

To check for per-game differences, we decomposed the ratings into a global component and game-specific deviations as in Equation \ref{elo2}. This means that we optimize six parameters for each player - one global rating and five per-game deviations.

To estimate the ratings, we made two kinds of pairwise comparisons between players: 

\begin{enumerate}
    \item their scores in games in which they appeared together,
    \item their respective scores when they played separately in "parallel games", i.e. different instances of identical game types and sizes against identical opponents.
\end{enumerate}

The latter is important since in cooperative situations, one player can be a spoiler that drags down everyone's scores. In this case, comparing the spoiler to a non-spoiler only in games where they played together would be unfair. But we could make a fair comparison between them by examining their parallel games. Parallel games are also beneficial for cases where two agents never or rarely faced each other in the same game.

\begin{figure}[h!]
    \centering
    \includegraphics[width=0.9\linewidth]{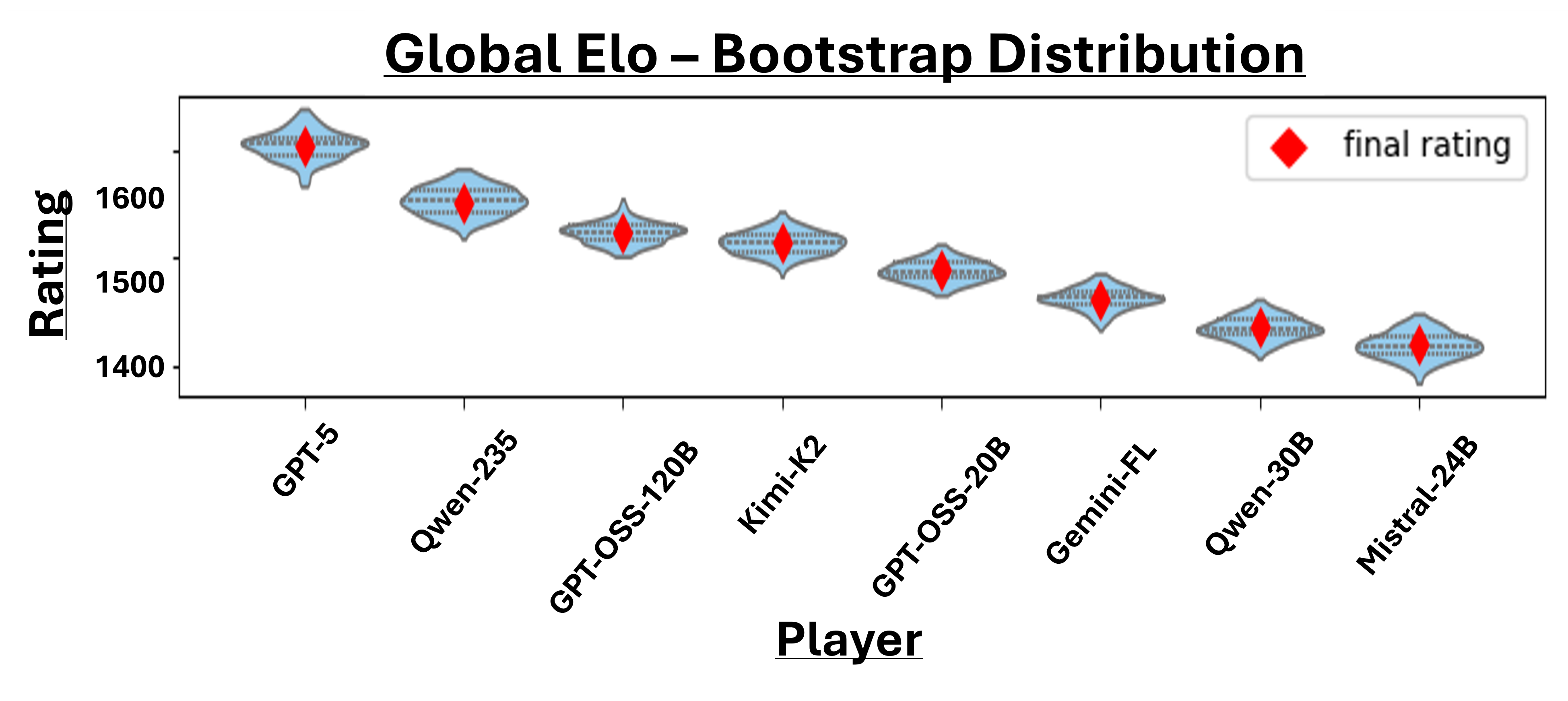}    
    \caption{Bootstrap test for global ELO ratings. Violins represent distribution of Elo scores across 200 subsampled re-estimations.}
    \label{violin}
\end{figure}

Results are presented in table \ref{elo-results}. For statistical significance, we ran a bootstrap test\cite{Bootstrap}, sub-sampling from the games 200 times and recalculating Elo ratings. Results of the bootstrap are visualized in figure \ref{violin}.

Testing the predictive power of the Elo model, we report an ROC-AUC of 0.67 when using only the global Elo ratings, a number that is significantly larger than expected by chance. The improvement when adding the game-specific biases to the ratings is small, 0.69, which indicates limited fluctuations in per-game rankings. This can be more directly quantified by the average Kendall-Tau correspondence of per-game rankings to the global ranking, which is 0.68. It can be concluded that there is an inherent social intelligence scalar, which non-trivially delineates the LLM agents' performance across social games. 

However, the model does not fully determine outcomes. For example, the lowest ranked agent outperformed the highest ranked agent about a quarter of the time. And there is nonetheless some fluctuation in agent rankings between games.

To test whether the limitations in the delineation of agent social intelligence is due to the constraint of estimating a single skill parameter with some per-game biases, we fitted the alternative model in Equation \ref{w_g}, estimating multiple latent per-player parameters with multiplicative per-game vector scalers, trying increasing dimensionality from 1 to 7. Each time we tested ROC-AUC using 5-fold cross-validation. The ROC-AUC for each dimension-size tried was in the range [0.67, 0.70]. The lack of substantial improvement over the Elo model indicates that we do not gain much from the more complex model of multiple agent skills each with different scalings in different games.

\section{RQ2: Socio-Cognitive Factors in LLM Agents}
\subsection{Metrics}
Beyond the scores at the end of each game, the gameplay data allows us to make a more fine-grained socio-cognitive investigation of the LLM-agents.
For this, we need to define a set of relevant metrics. The metrics chosen were: "Theory of Mind" (anticipation of others' actions), "predictability" (transparency of agents' intentions), "assertiveness" (choice of actions with immediate cost to others), "influence" (power of communication to affect reasoning of others), "amenability" (susceptibility to influence by others), "planning" (depth of look-ahead during reasoning) and "learning" (drawing lessons from previous rounds during reasoning). 

To evaluate "Theory of Mind", we measured the accuracy with which each player predicted the actions of each other player at each round of the game. Since each game has different action space, separate accuracy scores were designed. Note that each accuracy score is associated with a dyad of players, i.e. how player A predicted the actions of player B. By reversing the dyads, we obtain  "predictability" scores.

For "assertiveness", we examined the extent to which each player's decisions imposed costs on other players in each round, based on the structure of the actions and preferences of the game. In Survivor, we measured the proportion of available ammunition used by a player to attack others in a given turn. In Coalition and in Tragedy of the Commons, we measured the proportion of the available common resource demanded/extracted by a player in a given turn. In Scheduler, we measured the value of each proposal as a proportion of the highest individual preference of the proposer (higher preferences conflict more with the other players). In HUPI, we measured the proportion of the bid to the highest possible bid (as bids have a blocking effect on each other).

\begin{figure}
    \centering
    \includegraphics[width=0.9\linewidth]{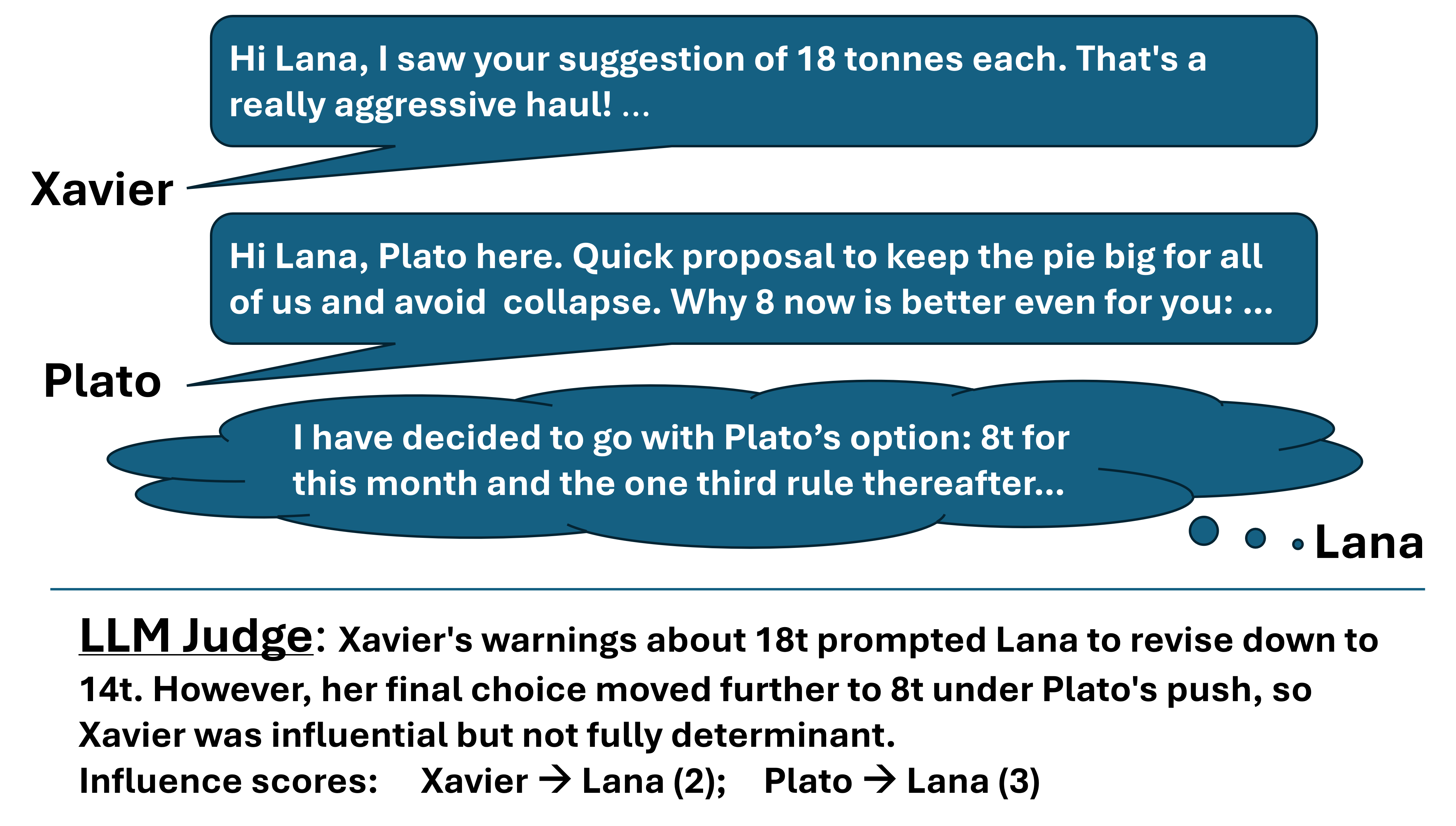}    
    \caption{Example illustrating how the LLM-judge rates influence in the Tragedy of Commons game where the players must decide how many tonnes to haul from a fishery, maximizing their individual gains while not collectively overfishing. The LLM judge examines the dialogues that a player had and the reasoning for their final decision at each turn.}
    \label{dialog}
\end{figure}

To evaluate the less tangible aspects, we used an LLM-judge (GPT-5 mini, reasoning set to high). To measure the "influence" that agents had over each other through communication, the LLM-judge was given the conversations that the players had, and the subsequent private reasoning and decisions of the players, in each round. It was prompted to rate the extent to which each pairwise conversation affected the reasoning and actions of each of the participants, on a scale of 0 to 3 (from no influence to decisive influence). An illustration is given in figure~\ref{dialog}. Once again, this produces a set of dyads. Reversing the dyad gives us the "amenability" score.

To rate the strategic abilities of the agents, the LLM-judge was fed the history of the observations, the private reasoning and the actions of a player through a game. It was prompted to rate the level of "planning" exhibited during reasoning at each round on a scale of 0 to 3 (from no/faulty planning to accurate multi-step planning). The same history was also used to prompt the LLM-judge to rate the "learning" exhibited during reasoning at each round, also on a scale of 0 to 3 (from inflexibility or drawing wrong lessons, to effective learning and adaptation).

The LLM-judge scores were validated by comparing to a different LLM-judge (Qwen3 Max 80B). Analyzing 25847 paired integer scores ranging from 0 to 3, the LLM judges had a mean absolute difference of 0.69 between their evaluations. Their scores matched precisely more than half the time (13961) and were within one point difference of each other over 80\% of the time (20939).

The socio-cognitive metrics are summarized in table \ref{metrics}.

\begin{table}
    \centering
    \begin{tabular}{ccccccc}
    \hline
    Metric           & Evaluation Method                        \\ 
    \hline
    Theory of Mind   & Accuracy of action predictions           \\
    Transparency     & Inverse of Theory of Mind                \\
    Influence        & LLM comparing chats to actions     \\
    Amenability      & Inverse of Influence                     \\
    Assertiveness    & Compare action values to preferences    \\
    Planning         & LLM rates planning depth in reasoning \\
    Learning         & LLM rates learning in reasoning \\
    Game Outcome     & Score by defined game logic              \\
    \hline
    \end{tabular}
    \caption{Socio-cognitive metrics}
        \label{metrics}
\end{table}

\subsection{Socio-Cognitive Consistency Results}

\begin{figure}
    \centering    
    \includegraphics[width=0.9\linewidth]{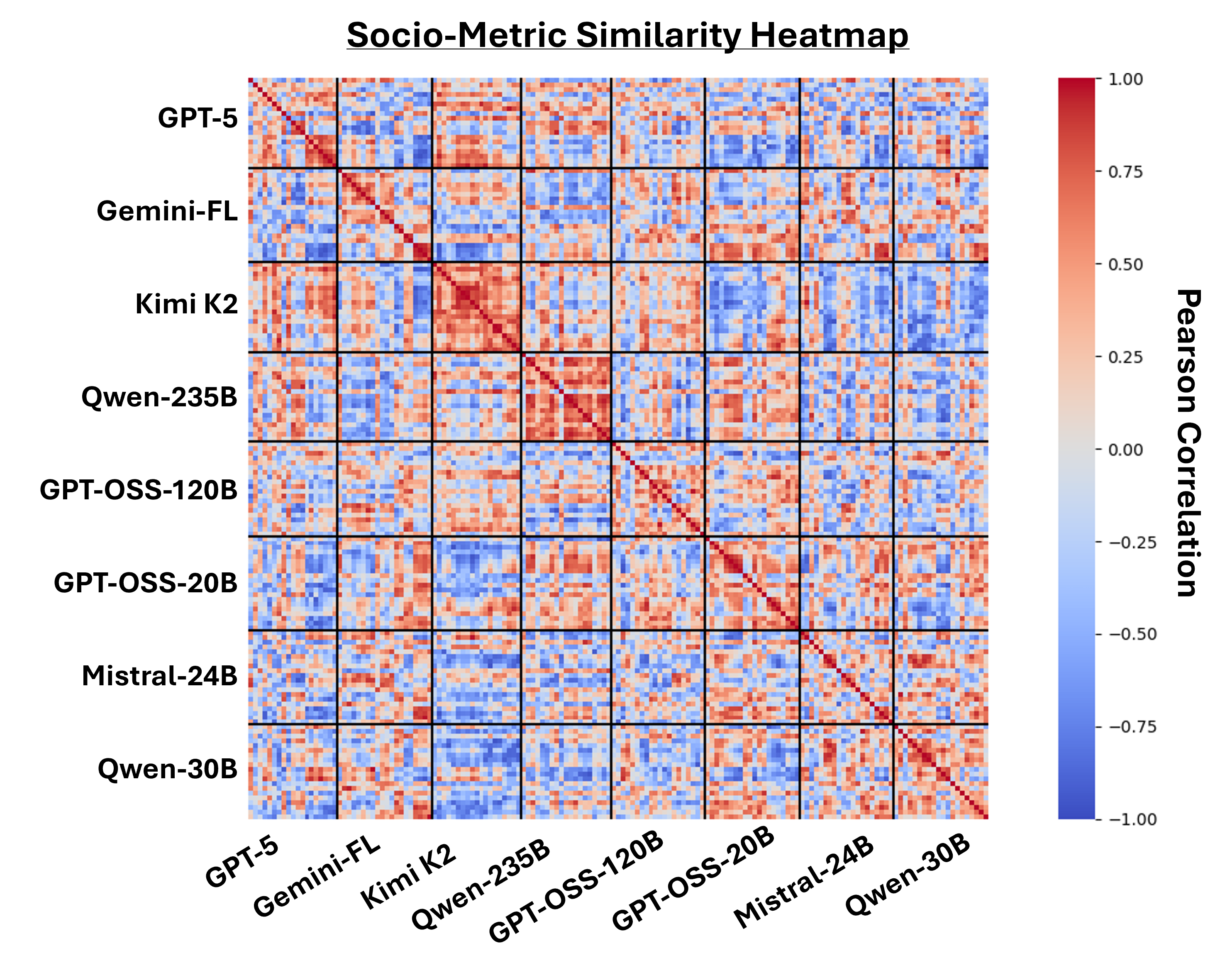}
    \caption{Socio-cognitive metrics were averaged within each game category for each player and then compared pairwise. Each large box represents the metrics of a player, and within it there are 20x20 smaller boxes representing the game categories. Colors represent Pearson correlations. The blocks on the diagonal (intra-player) show higher correlations than the off-diagonal. In aggregate, diagonal blocks correlate internally at 0.28 whereas off-diagonal regions correlate at 0.007.}
    \label{similarity heatmap}
\end{figure}

We examined the Pearson correlations between the vectors of socio-cognitive metrics gathered from different agents across the game matches. Averaging metrics of each agent within each of the 20 game categories (5 games x 4 sizes), we compared agent metrics to themselves across different game categories, and we compared the metrics of different agents. These Pearson correlations are visualized in the similarity heatmap in Figure \ref{similarity heatmap}.

To test whether players or games drive metric differences, we compared mean \textit{intra}-agent correlations across different games (0.27) to mean \textit{inter}-agent correlations in identical games (-0.05). Agents have much more internal consistency than consistency with others, regardless of the game. Some agents, such as Kimi-K2 and Qwen-235B demonstrated particularly-high self-consistency with intra-agent correlations of 0.56 and 0.48 respectively. 

\begin{figure}
    \centering    
    \includegraphics[width=0.9\linewidth]{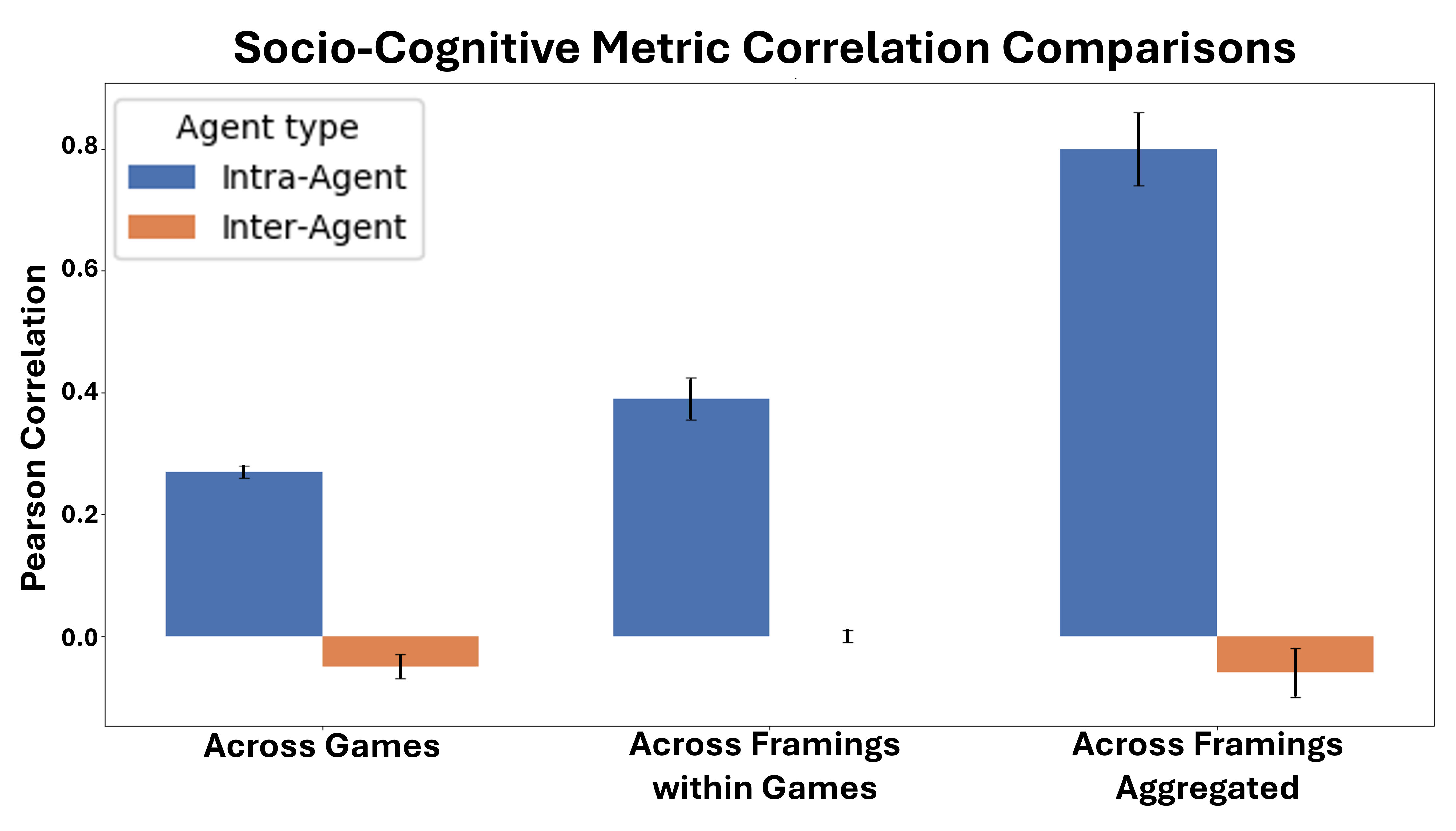}
    \caption{Comparison between intra-agent and inter-agent Pearson correlations of socio-cognitive metrics. Left is the comparison across different game types, middle is the comparison within game types across framings, right is comparisons across framings when averaging metrics in all games. Error bars represent Standard Error of the Mean.}
    \label{intra-inter}
\end{figure}

Recall that we ran two different semantic framings for each game type. To test for sensitivity to framing, we again aggregate the socio-cognitive metrics of each player in each game category, but now divided into the two framings. When comparing each player to itself across the two framings in each game category, we get a mean Pearson correlation of 0.39. When comparing between different players within the same framing we get -0.005.
If we aggregate not just at the game category level, but take the mean socio-cognitive metrics of each player across all games of all types within each of the two framings, and then compare just these two vectors, we get Pearson correlations ranging from 0.59 to 0.96, with a mean of 0.8 across all players. By contrast, if we compare the vectors of averaged socio-cognitive metrics between \textit{different} players within the same framing, we get a mean Pearson correlation of -0.06 across the different player comparisons. Intra-agent correlations rise as we aggregate and decrease noise, but inter-agent correlations remain near-zero. Variation in players is far more important than variation in framing for the realization of socio-cognitive features. Results are visualized in figure \ref{intra-inter}

\newpage

This gives us an upper bound on the noise induced by framing differences in prompts, and reaffirms the relative distinctiveness and internal consistency of the LLM-agents.

\section{RQ3: Determinants of Social Intelligence}
\subsection{Formalization of Social Intelligence Determinants}
The Elo model assumed that agents have fixed attributes that predict outcomes. A different view is to focus on the socio-cognitive metrics of each player per game match, regardless of who they are. While we saw that agents had a basic self-consistency, there was also some fluctuation. Therefore, these metrics might be more predictive of performance than the identity of the agents. 

Let us denote the vector of socio-cognitive metrics for player $i$ in game event $e$ as $\boldsymbol{m_{i,e}}$. We use logistic regression to estimate the model:

\begin{equation}
    \log[P_{i, j \sim{A}, e\sim{E}}(R_{i, e} > R_{j, e})]  \propto (\boldsymbol{m_{i,e}} - \boldsymbol{m_{j,e}})\cdot \boldsymbol{g}
    \label{m_h}
\end{equation}

Where $\boldsymbol{g}$ are the parameters that weight the different metrics. Alternatively, we can multiply by a per-game vector of parameters $\boldsymbol{g_k}$ as before, representing the relative importance of each metric per-game type. If such a model were to outperform the Elo model of a single fixed social skill factor, then it could give us insight into the relative importance of different cognitive functions or behaviors comprising social intelligence, as derived from the weights in $\boldsymbol{g}$ or $\boldsymbol{g_k}$.

\subsection{Results for Socio-Cognitive Determinants}
Fitting the model in equation \ref{m_h} with the empirically derived metrics in table \ref{metrics}, we evaluate using cross-validation. However, since we know that these metrics correlate with the identities of agents, we control for the possibility that the identity of the agents serve as a confounding variable. This is accomplished by defining our validation folds across the agents, leaving out two of the agents from the data in each training run, and then testing on the data that involves each of the the two left-out players. ROC-AUC of this model improved substantially over the previous models to a mean of 0.75 across the validation folds in the case of a fixed weights vector, and improved further to 0.82 in the case of per-game weights vectors. This substantial improvement shows us that social intelligence is decomposable into cognitive/behavioral factors, which also have variable importance across games.

\begin{figure}
    \centering    
    \includegraphics[width=0.9\linewidth]{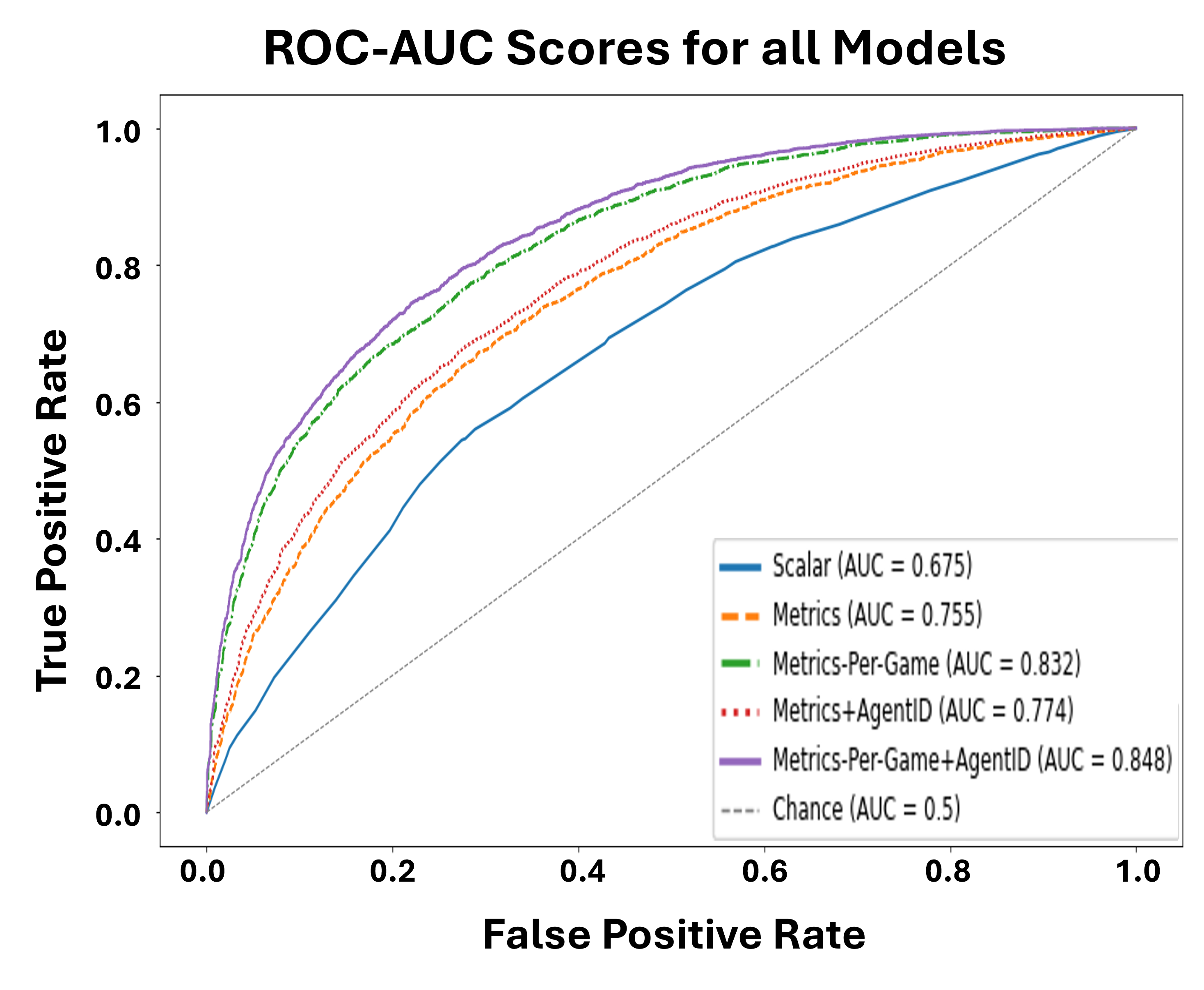}
    \caption{ROC-AUC curves for each model. All models are significantly above chance. Prediction via socio-cognitive metrics is substantially better than a single scalar. Per-game estimations are even better. Adding in agent identities as one-hots has an additional residual contribution. Results in the figure are for the trained models over all data.}
    \label{ROC-AUC}
\end{figure}

\begin{figure}
    \centering
    \begin{subfigure}{0.5\textwidth}
        \includegraphics[width=0.9\linewidth]{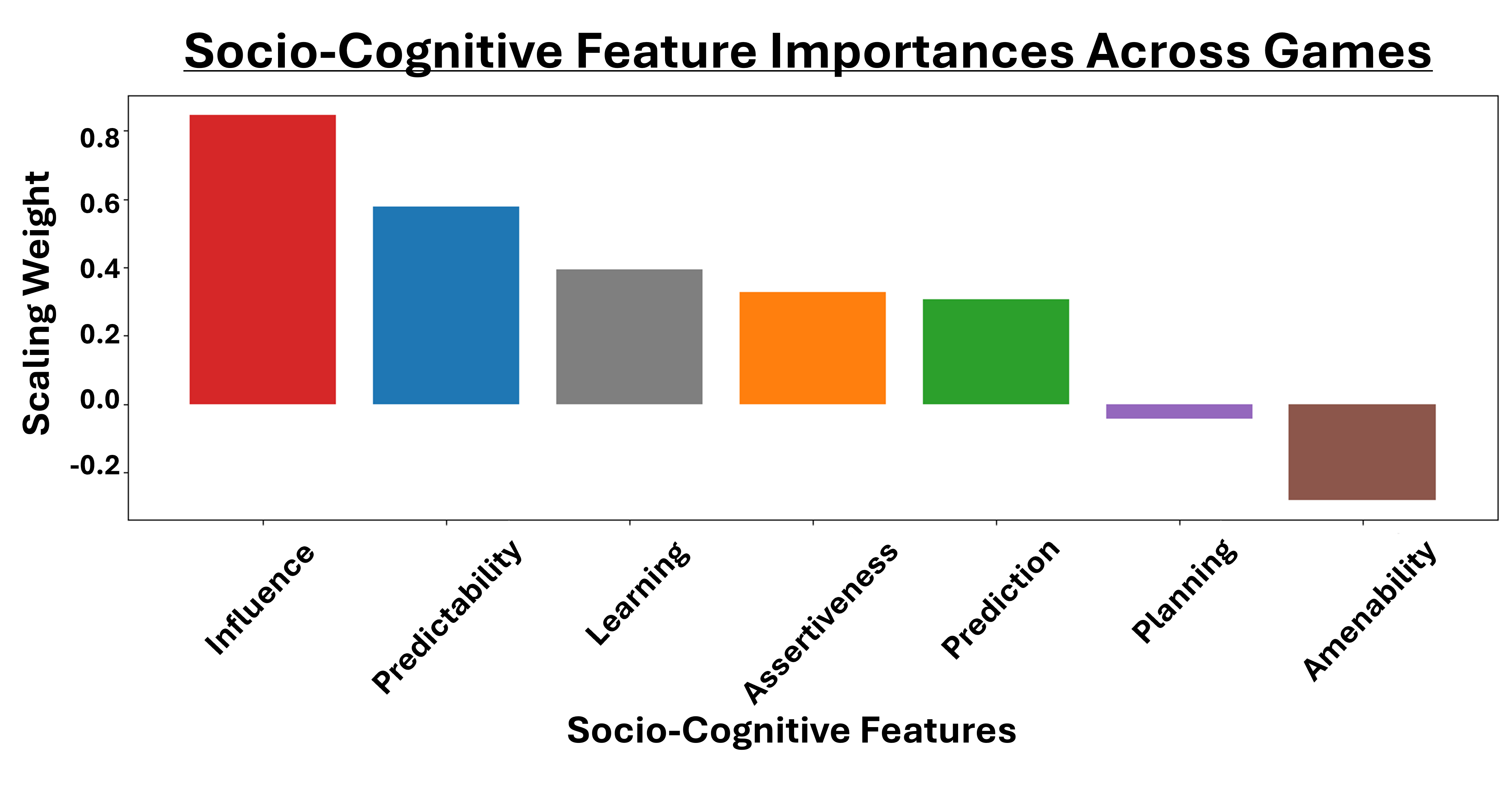}
        \caption{Fixed feature weights in model from Equation \ref{m_h}}
    \end{subfigure}
    \begin{subfigure}{0.5\textwidth}
        \includegraphics[width=0.9\linewidth]{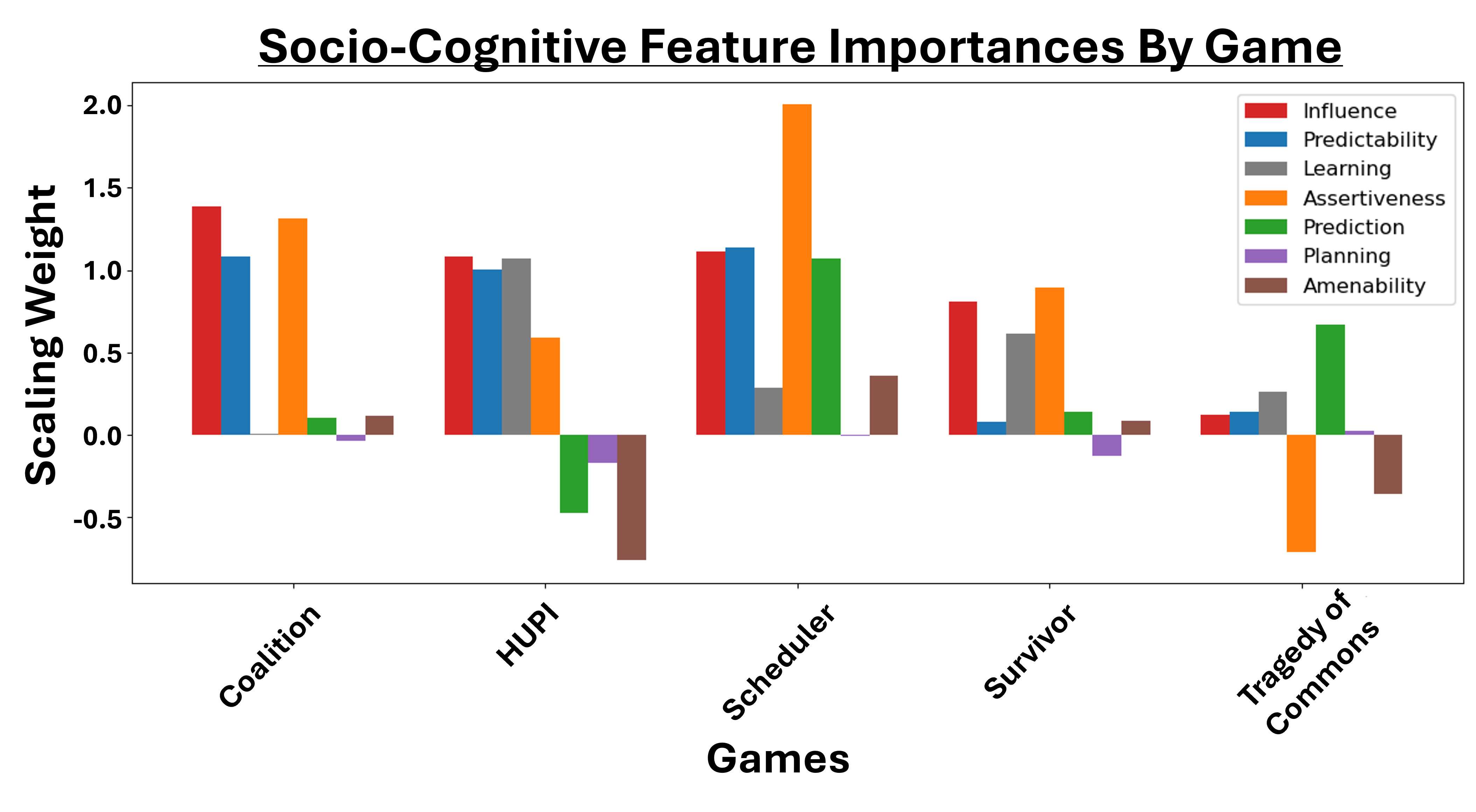}
        \caption{Feature weights per game in model from Equation \ref{m_h}}
    \end{subfigure}
    \caption{Feature Importances}
    \label{feature importances}
\end{figure}

Figure \ref{feature importances} illustrates the feature weights as estimated in the fixed $\boldsymbol{g}$ version of the model. They give us the somewhat surprising result that "influence" and "predictability" are the most important socio-cognitive metrics, followed by "learning" and "assertiveness". The least important features were "planning" and "prediction" ("Theory of Mind"). Perhaps unsurprisingly, "amenability" was negatively correlated with game outcomes. 

The per-game vector model can give some insight about the relative importance of socio-cognitive features in different games. For example, assertiveness stands out as being important in all games, but sometimes in a positive direction and sometimes in a negative direction, so its importance is obscured in the single vector model. This nonlinearity is probably a feature of mixed cooperative-competitive games. 

To test for residual individualized social intelligence, we fit a model that includes both the differences between the socio-cognitive metrics of the agents in a game \textit{and} their identities (using one-hot vectors). Evaluated with 4-fold cross-validation (this time randomly across matches, not across players, since we are using agent identities as inputs), we found that the game-invariant model achieves mean ROC-AUC of 0.77, and the game-sensitive model achieves 0.84. That is an improvement of 0.02 over the socio-cognitive metrics alone in both cases. The modest increase indicates that there is some residual agent-specific social intelligence, albeit small, that is not captured in the socio-cognitive metrics.

\section{Discussion}
\subsection{Conclusions}

We illustrated the central importance of communication in differentiating the social intelligence of LLM agents (see ablation study). Communication is an essential element in human interactions and the main training data of LLMs. The most important socio-cognitive features we identified were influence and transparency, which are intrinsically linked to communication. The complexity of natural language creates a particularly rich space for agents to maneuver in. Our work partially characterizes the intelligence required to succeed in that space.

We found that LLM agents have a scalar social intelligence factor which modestly differentiates between them. We also found that they have basically consistent socio-cognitive profiles, with some fluctuation. The manifestation of these socio-cognitive metrics per match are reliably predictive of game outcomes. The fluctuations in these metrics within agents across matches can partly account for the limited differentiation between them in the single social intelligence scalar.

Examining the socio-cognitive factors separately from the agents, we found that influence, transparency and learning were more important for social intelligence than "Theory of Mind" or planning\cite{dan2025behavior}. One interpretation might be: it is better to engage in cheap talk, actively influence events and be adaptive than to overthink the situation. 

\subsection{Previous Work}
Previous works have only partially or indirectly approached our research questions. For example, there exist static question-answering benchmarks for social intelligence, for humans\cite{SocialIntelligence} and LLMs\cite{SocialIQA}, but they do not test actual behavior in dynamic situations. There has also been work on measuring particular social cognitive faculties in humans and/or LLMs, such as "Theory of Mind"\cite{TestingTheoryOfMind}, persuasion\cite{Persuasion}, deception\cite{Deception} or bargaining\cite{Bargaining}, but it is unclear whether and how they relate to a general social intelligence. 
LLM agents were deployed in dynamic multiagent games in \cite{Sotopia,Concordia,GamaBench,StatesasStrings}.
\cite{Sotopia} defines a social intelligence metric by composing different indicators bottom-up. We adopt the unifying principle of reward maximization, and then employ a top-down approach to search for factors of social intelligence that can be justified by contribution to performance.
\cite{GamaBench} is a benchmark testing the ability of LLMs in games, but it does not feature inter-agent communication, and its scoring system is based on distance from the Nash Equilibrium best-response rather than pure reward.
\cite{StatesasStrings} introduced "chat games" but their focus was on the use of game-theoretic solvers to augment LLMs.
\cite{Concordia} is closest to our work. However, the authors focused exclusively on cooperation ability, and they did not directly measure socio-cognitive attributes of the agents' reasoning or behaviors during play, relying instead on subjective game attribute tagging as features.

\subsection{Contributions}
We propose a framework and concrete benchmark\footnote{Our game suite competed in the AgentX-AgentBeats Competition hosted by UC Berkeley Center for Responsible Distributed Intelligence and the Agentic AI MOOC, winning third place in the multiagent benchmark category, https://rdi.berkeley.edu/agentx-agentbeats.html} for the principled and systematic evaluation of social intelligence.
We demonstrate that current LLM-based agents are moderately differentiated by a single social intelligence factor, as exhibited by an Elo score. We also show that LLM-agents with similar prompts produce relatively self-consistent socio-cognitive profiles.
Lastly, we offer fresh empirical insights into the determinants of social intelligence per se, relating socio-cognitive profiles to performance.

\subsection{Limitations and Future Work}
We used LLM-judges and validated it with an independent LLM. The LLM-judge successfully picked up on signals predictive of final game outcomes given only localized information (e.g. "influence"), which constitutes additional validation. Nonetheless, we will seek additional human validation of this method. 
Our work is so far limited to LLM-agents. However, the framework is flexible enough to support human participation. Future work will expand the analysis to human-human and human-machine interactions. 
Our work has focused on evaluation. However, the framework is also a potential training environment. Future work will explore "social post-training" for LLM-agents. 

\subsection{Ethical Statement}
Ethical considerations play an important role in our motivations. Both socially inept and socially super-intelligent agents pose ethical risks. Our work takes a step in quantifying where LLM agents might be along this spectrum.

\subsection{Acknowledgments}
We would like to thank OpenAI for a generous contribution of tokens for LLM usage. This work was supported in part by the Israel Ministry of Innovation, Science and Technology grant 1001818511 (S.k.) and by the Gatsby Charitable Foundation (Y.L.). Y.L. is incumbent of the David and Inez Myers Chair in Neural Computation.

\bibliographystyle{named}
\bibliography{ijcai26}

@inproceedings{Concordia,
  title={Evaluating Generalization Capabilities of LLM-Based Agents in Mixed-Motive Scenarios Using Concordia},
  author={Chandler Smith and Marwa Abdulhai and Manfred Diaz and Marko Tesic and Rakshit Trivedi and Alexander Sasha Vezhnevets and Lewis Hammond and Jesse Clifton and Minsuk Chang and Edgar A. Du{\'e}{\~n}ez-Guzm{\'a}n and John P. Agapiou and Jayd Matyas and Danny Karmon and Akash Kundu and Aliaksei Korshuk and Ananya Ananya and Arrasy Rahman and Avinaash Anand Kulandaivel and Bain McHale and Beining Zhang and Buyantuev Alexander and Carlos Saith Rodriguez Rojas and Caroline Wang and Chetan Talele and Chenao Liu and Chichen Lin and Diana Riazi and Di Yang Shi and Emanuel Tewolde and Elizaveta Tennant and Fangwei Zhong and Fuyang Cui and Gang Zhao and Gema Parre{\~n}o Piqueras and Hyeonggeun Yun and Ilya Makarov and Jiaxun Cui and Jebish Purbey and J. E. Dilkes and Jord Nguyen and Lingyun Xiao and Luis Felipe Giraldo and Manuela Chacon-Chamorro and Manuel Sebastian Rios Beltran and Marta Emili Garc'ia Segura and Mengmeng Wang and Mogtaba Alim and Nicanor Quijano and Nico Schiavone and Olivia Macmillan-Scott and Oswaldo Pena and Peter Stone and Ram Mohan Rao Kadiyala and Rolando Fernandez and Rub{\'e}n Manrique and Su Lu and Sheila A. McIlraith and Shamika Dhuri and Shuqing Shi and Siddhant Gupta and Sneheel Sarangi and Sriram Ganapathi Subramanian and Taehun Cha and Toryn Q. Klassen and Wenming Tu and Weijian Fan and Ruiyang Wu and Xue Feng and Yali Du and Yang Liu and Yiding Wang and Yipeng Kang and Yoo Yeon Sung and Yuxuan Chen and Zhaowei Zhang and Zhihan Wang and Zhiqiang Wu and Ziang Chen and Zilong Zheng and Zixia Jia and Ziteng Wang and Dylan Hadfield-Menell and Natasha Jaques and Tim Baarslag and Jos{\'e} Hern{\'a}ndez-Orallo and Joel Z. Leibo},
  year={2025},
  url={https://api.semanticscholar.org/CorpusID:283247574}
}

@article{KrausDiplomacy,
  title={Designing and building a negotiating automated agent (Diplomacy)},
  author={Kraus, Sarit and Lehmann, Daniel},
  journal={Computational Intelligence},
  volume={11},
  number={1},
  pages={132--171},
  year={1995},
  publisher={Wiley Online Library}
}

@inproceedings{GamaBench,
  author    = {Jen{-}tse Huang and
               Eric John Li and
               Man Ho Lam and
               Tian Liang and
               Wenxuan Wang and
               Youliang Yuan and
               Wenxiang Jiao and
               Xing Wang and
               Zhaopeng Tu and
               Michael R. Lyu},
  title     = {Competing Large Language Models in Multi-Agent Gaming Environments},
  booktitle = {Proceedings of the Thirteenth International Conference on Learning Representations (ICLR)},
  year      = {2025}
}

@inbook{SocialIntelligence, place={Cambridge}, series={Cambridge Handbooks in Psychology}, title={Social Intelligence}, booktitle={The Cambridge Handbook of Intelligence}, publisher={Cambridge University Press}, author={Kihlstrom, John F. and Cantor, Nancy}, year={2011}, pages={564–581}, collection={Cambridge Handbooks in Psychology}}

@inproceedings{SocialIQA,
    title = "Social {IQ}a: Commonsense Reasoning about Social Interactions",
    author = "Sap, Maarten  and
      Rashkin, Hannah  and
      Chen, Derek  and
      Le Bras, Ronan  and
      Choi, Yejin",
    editor = "Inui, Kentaro  and
      Jiang, Jing  and
      Ng, Vincent  and
      Wan, Xiaojun",
    booktitle = "Proceedings of the 2019 Conference on Empirical Methods in Natural Language Processing and the 9th International Joint Conference on Natural Language Processing (EMNLP-IJCNLP)",
        year = "2019",
    address = "Hong Kong, China",
    publisher = "Association for Computational Linguistics",
        pages = "4463--4473",
   }

@inproceedings{zuckerman2009mixing,
  title={Mixing Search Strategies for Multi-Player Games.},
  author={Zuckerman, Inon and Felner, Ariel and Kraus, Sarit},
  booktitle={IJCAI},
  volume={9},
  pages={646--652},
  year={2009}
}

@article{TestingTheoryOfMind, title={Testing theory of mind in large language models and humans}, volume={8}, DOI={10.1038/s41562-024-01882-z}, number={7}, journal={Nature Human Behaviour}, author={Strachan, James W. and Albergo, Dalila and Borghini, Giulia and Pansardi, Oriana and Scaliti, Eugenio and Gupta, Saurabh and Saxena, Krati and Rufo, Alessandro and Panzeri, Stefano and Manzi, Guido and et al.}, year={2024}, month={May}, pages={1285–1295}}

@inproceedings{Persuasion,
    title = "Measuring and Benchmarking Large Language Models' Capabilities to Generate Persuasive Language",
    author = "Pauli, Amalie Brogaard  and
      Augenstein, Isabelle  and
      Assent, Ira",
    editor = "Chiruzzo, Luis  and
      Ritter, Alan  and
      Wang, Lu",
    booktitle = "Proceedings of the 2025 Conference of the Nations of the Americas Chapter of the Association for Computational Linguistics: Human Language Technologies (Volume 1: Long Papers)",
    month = apr,
    year = "2025",
    address = "Albuquerque, New Mexico",
    publisher = "Association for Computational Linguistics",
    url = "https://aclanthology.org/2025.naacl-long.506/",
    doi = "10.18653/v1/2025.naacl-long.506",
    pages = "10056--10075",
    ISBN = "979-8-89176-189-6"
}

@article{Deception, title={Deception abilities emerged in large language models}, author={Hagendorff, Thilo}, journal={Proceedings of the National Academy of Sciences}, volume={121}, number={24}, year={2024}, doi={10.1073/pnas.2317967121}, url={https://doi.org/10.1073/pnas.2317967121} }

@inproceedings{Bargaining, author = {Deuksin Kwon and Emily Weiss and Tara Kulshrestha and Kushal Chawla and Gale Lucas and Jonathan Gratch}, title = {Are LLMs Effective Negotiators? Systematic Evaluation of the Multifaceted Capabilities of LLMs in Negotiation Dialogues}, booktitle = {Findings of the Association for Computational Linguistics: EMNLP 2024}, pages = {5391--5413}, year = {2024}, address = {Miami, Florida, USA}, publisher = {Association for Computational Linguistics} }

@misc{Sotopia,
      title={SOTOPIA: Interactive Evaluation for Social Intelligence in Language Agents}, 
      author={Xuhui Zhou and Hao Zhu and Leena Mathur and Ruohong Zhang and Haofei Yu and Zhengyang Qi and Louis-Philippe Morency and Yonatan Bisk and Daniel Fried and Graham Neubig and Maarten Sap},
      year={2024},
      eprint={2310.11667},
      archivePrefix={arXiv},
      primaryClass={cs.AI},
      url={https://arxiv.org/abs/2310.11667}, 
}

@article{RewardisEnough,
title = {Reward is enough},
journal = {Artificial Intelligence},
volume = {299},
pages = {103535},
year = {2021},
issn = {0004-3702},
doi = {https://doi.org/10.1016/j.artint.2021.103535},
url = {https://www.sciencedirect.com/science/article/pii/S0004370221000862},
author = {David Silver and Satinder Singh and Doina Precup and Richard S. Sutton}
}

@article{Diplomacy,
author = {Anton Bakhtin  and Noam Brown  and Emily Dinan  and Gabriele Farina  and Colin Flaherty  and Daniel Fried  and Andrew Goff  and Jonathan Gray  and Hengyuan Hu  and Athul Paul Jacob  and Mojtaba Komeili  and Karthik Konath  and Minae Kwon  and Adam Lerer  and Mike Lewis  and Alexander H. Miller  and Sasha Mitts  and Adithya Renduchintala  and Stephen Roller  and Dirk Rowe  and Weiyan Shi  and Joe Spisak  and Alexander Wei  and David Wu  and Hugh Zhang  and Markus Zijlstra },
title = {Human-level play in the game of <i>Diplomacy</i> by combining language models with strategic reasoning},
journal = {Science},
volume = {378},
number = {6624},
pages = {1067-1074},
year = {2022},
doi = {10.1126/science.ade9097},
URL = {https://www.science.org/doi/abs/10.1126/science.ade9097},
eprint = {https://www.science.org/doi/pdf/10.1126/science.ade9097}}

@Inbook{Coalition,
author="Leyton-Brown, Kevin
and Shoham, Yoav",
title="Coalitional Game Theory",
bookTitle="Essentials of Game Theory: A Concise, Multidisciplinary Introduction",
year="2008",
publisher="Springer International Publishing",
address="Cham",
pages="69--77",
isbn="978-3-031-01545-8",
doi="10.1007/978-3-031-01545-8_8",
url="https://doi.org/10.1007/978-3-031-01545-8_8"
}

@article{StatesasStrings,
  author       = {Ian Gemp and
                  Yoram Bachrach and
                  Marc Lanctot and
                  Roma Patel and
                  Vibhavari Dasagi and
                  Luke Marris and
                  Georgios Piliouras and
                  Siqi Liu and
                  Karl Tuyls},
  title        = {States as Strings as Strategies: Steering Language Models with Game-Theoretic Solvers},
  journal      = {CoRR},
  volume       = {abs/2402.01704},
  year         = {2024},
  url          = {https://doi.org/10.48550/arXiv.2402.01704},
  doi          = {10.48550/ARXIV.2402.01704},
  eprinttype    = {arXiv},
  eprint       = {2402.01704},
  bibsource    = {dblp computer science bibliography, https://dblp.org}
}

@inproceedings{TragedyofCommons,
 author = {Piatti, Giorgio and Jin, Zhijing and Kleiman-Weiner, Max and Sch\"{o}lkopf, Bernhard and Sachan, Mrinmaya and Mihalcea, Rada},
 booktitle = {Advances in Neural Information Processing Systems},
 doi = {10.52202/079017-3548},
 editor = {A. Globerson and L. Mackey and D. Belgrave and A. Fan and U. Paquet and J. Tomczak and C. Zhang},
 pages = {111715--111759},
 publisher = {Curran Associates, Inc.},
 title = {Cooperate or Collapse:  Emergence of Sustainable Cooperation in a Society of LLM Agents},
 url = {https://proceedings.neurips.cc/paper_files/paper/2024/file/ca9567d8ef6b2ea2da0d7eed57b933ee-Paper-Conference.pdf},
 volume = {37},
 year = {2024}
}

@article{Lupi,
Author = {Östling, Robert and Wang, Joseph Tao-yi and Chou, Eileen Y. and Camerer, Colin F.},
Title = {Testing Game Theory in the Field: Swedish LUPI Lottery Games},
Journal = {American Economic Journal: Microeconomics},
Volume = {3},
Number = {3},
Year = {2011},
Month = {August},
Pages = {1–33},
DOI = {10.1257/mic.3.3.1},
URL = {https://www.aeaweb.org/articles?id=10.1257/mic.3.3.1}}

@inbook{Axelrod,
  address = {Stanford},
  author = {Axelrod, Robert},
  booktitle = {Theories Of Social Order: A Reader},
  key = {axelrod:1984},
  location = {EA Soz:ES7},
  publisher = {Stanford University Press},
  title = {The Evolution Of Cooperation},
  year = 1984
}

@article{SocialCognition,
   author = "Happé, Francesca and Cook, Jennifer L. and Bird, Geoffrey",
   title = "The Structure of Social Cognition: In(ter)dependence of Sociocognitive Processes", 
   journal= "Annual Review of Psychology",
   year = "2017",
   volume = "68",
   number = "Volume 68, 2017",
   pages = "243-267",
   doi = "https://doi.org/10.1146/annurev-psych-010416-044046",
   url = "https://www.annualreviews.org/content/journals/10.1146/annurev-psych-010416-044046",
   publisher = "Annual Reviews",
   issn = "1545-2085",
   type = "Journal Article",
  }

@book{Influence, 
place={New York}, 
title={Influence: The psychology of persuasion}, 
publisher={HarperCollins}, 
author={Cialdini, Robert B.}, year={2021}}

@incollection{StrategicReasoning,
title = {Chapter One - Strategic Thinking},
editor = {James M. Olson and Mark P. Zanna},
series = {Advances in Experimental Social Psychology},
publisher = {Academic Press},
volume = {54},
pages = {1-66},
year = {2016},
issn = {0065-2601},
doi = {https://doi.org/10.1016/bs.aesp.2016.03.001},
url = {https://www.sciencedirect.com/science/article/pii/S006526011630017X},
author = {N. Halevy}
}

@article{Susceptibility,
title = {Social influence research in consumer behavior: What we learned and what we need to learn? – A hybrid systematic literature review},
journal = {Journal of Business Research},
volume = {162},
pages = {113870},
year = {2023},
issn = {0148-2963},
doi = {https://doi.org/10.1016/j.jbusres.2023.113870},
url = {https://www.sciencedirect.com/science/article/pii/S014829632300228X},
author = {Ramulu Bhukya and Justin Paul}
}

@incollection{SocialSkills,
title = {Chapter 1 - Defining social skills},
editor = {Douglas W. Nangle and Cynthia A. Erdley and Rebecca A. Schwartz-Mette},
booktitle = {Social Skills Across the Life Span},
publisher = {Academic Press},
pages = {3-24},
year = {2020},
isbn = {978-0-12-817752-5},
doi = {https://doi.org/10.1016/B978-0-12-817752-5.00001-9},
url = {https://www.sciencedirect.com/science/article/pii/B9780128177525000019},
author = {Rachel L. Grover and Douglas W. Nangle and Michelle Buffie and Laura A. Andrews}
}

@book{Communication,
place={London ; New York}, 
edition={7}, 
title={Skilled interpersonal communication : research, theory and practice}, ISBN={9781138823778}, 
publisher={Routledge, An Imprint Of The Taylor Et Francis Group}, 
author={Hargie, Owen}, 
year={2022} }

@article{Conflict,
author = {Rahim, M.},
year = {1983},
month = {10},
pages = {189-199},
title = {Measurement of Organizational Conflict},
volume = {109},
journal = {Journal of General Psychology},
doi = {10.1080/00221309.1983.10736085}
}

@book{SocialLearning, 
address={Englewood Cliffs, N.J.}, 
title={Social Learning Theory},
ISBN={9780138167448}, 
publisher={Prentice-Hall}, 
author={Bandura, Albert}, 
year={1977} }

@book{Elo, 
title={The Rating of Chessplayers, Past and Present}, publisher={B. T. Batsford Limited}, author={Elo, Arpad E}, year={1978} }

@article{Bootstrap,
   author = "Horowitz, Joel L.",
   title = "Bootstrap Methods in Econometrics", 
   journal= "Annual Review of Economics",
   year = "2019",
   volume = "11",
   number = "Volume 11, 2019",
   pages = "193-224",
   doi = "https://doi.org/10.1146/annurev-economics-080218-025651",
   url = "https://www.annualreviews.org/content/journals/10.1146/annurev-economics-080218-025651",
   publisher = "Annual Reviews",
   issn = "1941-1391",
   type = "Journal Article",
  }

@inproceedings{BradleyTerry,
  title={The many routes to the ubiquitous Bradley-Terry model},
  author={Ian Hamilton and Nick Tawn and David Firth},
  year={2023},
  url={https://api.semanticscholar.org/CorpusID:266435815}
}

@book{MLE, title={Mathematical Statistics}, 
ISBN={9781118770979}, 
publisher={John Wiley \& Sons}, 
author={Rossi, Richard J}, 
year={2018}, 
month={Jun} }

@article{SocialGames, title={The theory of social games: outline of a general theory for the social sciences}, volume={10}, DOI={https://doi.org/10.1057/s41599-023-01862-0}, number={1}, journal={Humanities \& social sciences communications}, publisher={Springer Nature}, author={Stolz, Jörg}, year={2023}, month={Jun} }

@article{Strategic-Communication, title={Strategic Information Transmission}, volume={50}, DOI={https://doi.org/10.2307/1913390}, number={6}, journal={Econometrica}, author={Crawford, Vincent P. and Sobel, Joel}, year={1982}, month={Nov}, pages={1431} }

@article{GroupSize,
author = {Avdiaj, Besnik},
year = {2022},
month = {03},
pages = {14-32},
title = {Size and Decision-making: A Systematic Literature Review on Groups and Teams},
volume = {7},
journal = {MANAGEMENT AND ECONOMICS REVIEW},
doi = {10.24818/mer/2022.02-02}
}

@article{BayesianGames, title={Games with Incomplete Information Played by {Bayesian} Players, I–III: Part I. The Basic Model}, volume={50},  journal={Management Science}, author={Harsanyi, John C.}, year={2004}, month={Dec}, pages={1804–1817} }

@inbook{MixedMotive, place={Cambridge}, title={Two-person mixed-motive games of strategy}, booktitle={Decision Making Using Game Theory: An Introduction for Managers}, publisher={Cambridge University Press}, author={Kelly, Anthony}, year={2003}, pages={98–134}}

@article{CognitiveHierarchy,
    author = {Camerer, Colin F. and Ho, Teck-Hua and Chong, Juin-Kuan},
    title = {A Cognitive Hierarchy Model of Games*},
    journal = {The Quarterly Journal of Economics},
    volume = {119},
    number = {3},
    pages = {861-898},
    year = {2004},
    month = {08},
    issn = {0033-5533},
    doi = {10.1162/0033553041502225},
    url = {https://doi.org/10.1162/0033553041502225},
    eprint = {https://academic.oup.com/qje/article-pdf/119/3/861/5461769/119-3-861.pdf},
}

@article{dan2025behavior,
  title={Behavior engineering using quantitative reinforcement learning models},
  author={Dan, Ohad and Plonsky, Ori and Loewenstein, Yonatan},
  journal={Nature Communications},
  volume={16},
  number={1},
  pages={4109},
  year={2025},
  publisher={Nature Publishing Group UK London}
}

\end{document}